# HIDDEN MARKOV MODELS FOR THE ASSESSMENT OF CHROMOSOMAL ALTERATIONS USING HIGH-THROUGHPUT SNP ARRAYS

By Robert B. Scharpf,[1] Giovanni Parmigiani,[2] Jonathan Pevsner[3] and Ingo Ruczinski[4]

*Johns Hopkins Bloomberg School of Public Health, Sidney Kimmel Comprehensive Cancer Center, Kennedy Krieger Institute and Johns Hopkins Bloomberg School of Public Health*

Chromosomal DNA is characterized by variation between individuals at the level of entire chromosomes (e.g., aneuploidy in which the chromosome copy number is altered), segmental changes (including insertions, deletions, inversions, and translocations), and changes to small genomic regions (including single nucleotide polymorphisms). A variety of alterations that occur in chromosomal DNA, many of which can be detected using high density single nucleotide polymorphism (SNP) microarrays, are linked to normal variation as well as disease and are therefore of particular interest. These include changes in copy number (deletions and duplications) and genotype (e.g., the occurrence of regions of homozygosity). Hidden Markov models (HMM) are particularly useful for detecting such alterations, modeling the spatial dependence between neighboring SNPs. Here, we improve previous approaches that utilize HMM frameworks for inference in high throughput SNP arrays by integrating copy number, genotype calls, and the corresponding measures of uncertainty when available. Using simulated and experimental data, we, in particular, demonstrate how confidence scores control smoothing in a probabilistic framework. Software for fitting HMMs to SNP array data is available in the R package *VanillaICE*.

Received July 2007; revised November 2007.

[1]Supported by NSF Grant DMS-03-4211 and training Grant 5T32HL007024 from the National Heart, Lung and Blood Institute.

[2]Supported by NSF Grant DMS-03-4211.

[3]Supported by NIH Grants HD046598 and HD24061.

[4]Supported by NIH Grants R01 CA074841 and R01 GM083084.

**1. Introduction.** Chromosomal DNA is characterized by variation between individuals at the level of entire chromosomes (e.g., aneuploidy in which the chromosome copy number is altered), segmental changes (including insertions, deletions, inversions, and translocations), and changes to small genomic regions (including single nucleotide polymorphisms). A variety of alterations that occur in chromosomal DNA, many of which can be detected using high density single nucleotide polymorphism (SNP) microarrays, are linked to normal variation as well as disease and are therefore of particular interest [Shaw-Smith et al. (2004), Aguirre et al. (2004), Aggarwal et al. (2005), Dutt and Beroukhim (2007), Sebat et al. (2007), Szatmari et al. (2007)]. These include changes in copy number (deletions and duplications) and genotype (e.g., the occurrence of regions of homozygosity).

Copy number variations can arise through somatic and germline events. While naturally occurring and often (but not always) benign, germline copy number variations are more abundant than previously thought [Freeman et al. (2006), Redon et al. (2006), Eichler et al. (2007)]. On the other hand, somatic copy number changes, such as gene amplifications and deletions, frequently contribute to tumorigenesis (or might be the consequence of it). Regions of homozygosity (i.e., long stretches of homozygous SNPs) can also occur through somatic and germline events. A hemizygous deletion of one chromosomal allele results in only one DNA copy, and therefore, SNPs in that region will appear as homozygous (given current genotyping technologies that generate only biallelic calls). The definition of loss of heterozygosity (LOH) refers to such a somatic event: for example, comparing a tumor and normal sample from the same person, any heterozygous SNPs in the normal sample appear as homozygous SNPs in the tumor sample, in any region where an allele was lost. As already noted, regions of homozygosity can also occur through germline events. While chromosomal DNA is typically inherited from both parents, under some circumstances an individual inherits two copies of a chromosome from one parent. The inheritance of both homologues of a pair of chromosomes from only one parent can be due to autozygosity (homozygosity in which alleles are identical by descent) or to uniparental disomy [UPD, Robinson (2000), Engel (2006)]. Autozygosity and UPD do not involve an aneuploidy (change in chromosomal copy number), and the region of homozygosity may extend over an entire chromosome or segmentally across a subregion of a chromosome. The condition is termed uniparental isodisomy (iUPD) if the two copies inherited from one parent are identical, and results in stretches of homozygous SNPs. (If the two inherited copies are different homologues, the result is uniparental heterodisomy, hUPD, but does not result in stretches of homozygous SNPs.) In some cases, UPD is thought to be benign, but can also be associated with disease [Prader–Willi syndrome, Angelman syndrome, Beckwith–Wiedemann syndrome, see, e.g.,

Altug-Teber et al. (2005)]. UPD can disrupt genomic imprinting, such that imprinted genes (expressed preferentially from the paternal or maternal alleles) fail to be expressed. UPD can also cause homozygosity for autosomal recessive traits such as cystic fibrosis [Zlotogora (2004)].

A variety of technologies have been applied for the assessment of chromosomal abnormalities including conventional karyotyping (e.g., Giemsa staining of metaphase chromosomes) and fluorescence in situ hybridization (FISH). While the former only allows for the genome-wide detection of major chromosomal amplifications and deletions, the latter allows for the verification of suspected microdeletions as well as translocations and some duplications. Array comparative genome hybridization (aCGH) permits a genome-wide measurement of copy number variation using bacterial artificial chromosome (BAC) clones deposited on a microarray. This is a high throughput technique, but the resolution is limited to tens or hundreds of thousands of base pairs and no genotype data are obtained.

SNP microarray technology permits the genome-wide search for chromosomal abnormalities, providing genotype and copy number estimates for hundreds of thousands of SNPs in genomic DNA isolated from a biological sample. Statistical tools for the analysis of such SNP chip data are typically employed to assess where the chromosomal changes have occurred, and whether or not these changes are associated with disease. Regions of interest are typically aneuploidies, that is, regions where copy number changes (deletions and amplifications) have occurred, or regions with unusually long stretches of homozygous genotypes (either naturally occurring, e.g., through evolutionary pressure on a DNA segment, or through loss of heterozygosity, LOH).

For the analysis of SNP chip data in general, three different tiers of estimation problems arise. (1) By SNP: how can we use the low-level data (such as the fluorescence measurements in Affymetrix SNP chips) to optimally estimate the genotype and DNA copy number for each SNP in the array? (2) By sample: how can we borrow strength between neighboring SNPs, and infer regions of LOH and copy number changes in the genome of the subject studied? (3) Between samples: how can we compare the genotype of many subjects, infer common regions of abnormality, and, for example, assess differences between affected subjects and normal controls? This manuscript revolves around methods for tier 2, the assessment of chromosomal abnormalities in one particular sample. However, information derived from tier 1, in particular, uncertainty estimates of copy number and genotype estimates, can be critically important and will be incorporated in the analysis. In particular, for the Affymetrix platform, originally described as a high-throughput assay for calling genotypes at thousands of SNPs [Kennedy et al. (2003)], there have been several algorithms proposed for the appropriate adjustment and pre-processing of probe-level data, and the estimation of SNP-level summaries of probe-level data for geno-
type [DM, Di et al. (2005), RLMM, Rabbee and Speed (2006), BRLMM,
Affymetrix (2006), CRLMM, Carvalho, Speed and Irizarry (2007), SNiPer-
HD, Hua et al. (2007)] and copy number (CNAG, Nannya et al. (2005),
CARAT, Huang et al. (2006), PLASQ, Laframboise, Harrington and Weir
(2007), CN-RLMM, Wang et al. (2007)]. Notably, Laframboise, Harrington
and Weir (2007) and Wang et al. (2007) provide allele-specific estimates of
copy number.

We caution that, as with gene expression technologies, pre-processing of
probe-level data is an important consideration. For instance, several recent
papers have described fragment-length and sequence effects that may be
introduced by the polymerase chain reaction (PCR) used to amplify the
DNA [Nannya et al. (2005), Carvalho, Speed and Irizarry (2007)]. We as-
sume that SNP-level summaries for each interrogated SNP have been ad-
justed for probe-specific biases to the extent possible. Statistical models such
as CRLMM that use Hapmap data for training have been shown to provide
better genotype calls when the centers of the bivariate scatterplots for the
$A$ and $B$ allele intensities are less well defined [Carvalho, Speed and Irizarry
(2007)]. Genotype calls for most genotyping algorithms are concordant for
over 99.9% of the measured SNPs in the Affymetrix 100k and 500k chips
when performance is compared on apparently normal individuals represented
in the HapMap study.

Statistical methods that provide an indication of the uncertainty of the
genotype call [e.g., based on the single to noise ratio (SNR) and log likelihood
ratio (LLR) defined by CRLMM] can be particularly useful for statistical
algorithms devised to infer chromosomal abnormalities. Specifically, statisti-
cal models that borrow strength from neighboring SNPs to infer loss or
retention of heterozygosity should incorporate the uncertainty of the geno-
type call estimate, giving less weight to genotype calls that are measured
with high uncertainty and more weight to well-estimated genotypes. To our
knowledge, this manuscript is the first one to address this issue. Figure 1
illustrates why the uncertainty in genotype calls can differ substantially. Sim-
ilarly, probe-specific biases for copy number estimates have been described
before, see, for example, Wang et al. (2007).

Before high-throughput SNP chips were widely available, array compara-
tive genomic hybridization (aCGH) was the most commonly used method to
assess DNA copy numbers, and assess regions in the genome where deletions
or amplifications occurred in a particular sample. Thus, many statisticians
have proposed approaches for aCGH based copy number estimation, and
some of these proposed methods are also relevant for SNP chip based copy
number analysis. Approaches for aCGH data include hidden Markov models
[Fridlyand et al. (2004), Guha, Li and Neuberg (2006)], segmentation algo-
rithms [Olshen et al. (2004), Picard et al. (2005), Venkatraman and Olshen

(2007)], wavelets [Hsu et al. (2005)], smoothing [Hupe et al. (2004), Eilers and de Menezes (2005)] regression [Houseman, Coull and Betensky (2006), Huang et al. (2005)], clustering [Wang et al. (2005)], and resampling [Lai and Zhao (2005)]. The manuscript by Lai et al. (2005) and Willenbrock and Fridlyand (2005) contain reviews and comparisons of the performances of several of these proposed methods. In addition, many useful extensions or alternative approaches for the above listed methods are being proposed. Some recent publications have confirmed that naturally occurring DNA copy number variations are more abundant than previously thought [Freeman et al. (2006), Redon et al. (2006)], which can produce outliers in the aCGH data. Integrating these known copy number variations as permissible outliers into a hidden Markov model to assess where abnormal copy number alterations have occurred has been proposed by Shah et al. (2006).

For the statistical analysis, SNP chip data differ from array CGH data in two important ways: (a) SNP chips also provide information for the genotype, that is, give homozygous/heterozygous SNP calls, and (b) provide a much denser coverage, currently generating genotype information and copy number estimates at locations in excess of 500,000 SNPs. The correlation structure between those estimates has to be an essential part of any statistical modeling approach. The most promising methods currently available are based on hidden Markov models. In particular, to infer LOH regions and to estimate copy numbers changes, the dChip software and methods are among the most widely used in the scientific literature for the analysis of SNP chip data. The dChip methods are based on separate hidden Markov Models for genotype analysis [Lin et al. (2004), Beroukhim et al. (2006)] and copy number [Zhao et al. (2004)]. The original dChipSNP HMM [Lin et al. (2004)] was devised to assess loss of heterozygosity regions (a region with an allelic loss, where heterozygote SNPs in a normal sample appear as homozygote SNPs in a tumor sample). This required paired tumor and normal samples from the *same subject*. As these are often not available, an extension of this model was proposed by Beroukhim et al. (2006) to allow for LOH assessment without paired samples (e.g., tumor only). Note that such an approach using unpaired data would also be required in settings that do not involve abnormal tissue, for example, when subjects with mental retardation and apparently normal controls are investigated to assess possible differences in the karyotypes. The dChipSNP hidden Markov Model for copy number assessment [Zhao et al. (2004)] is somewhat similar in nature to the one used for LOH analysis; see Zhao et al. (2004).

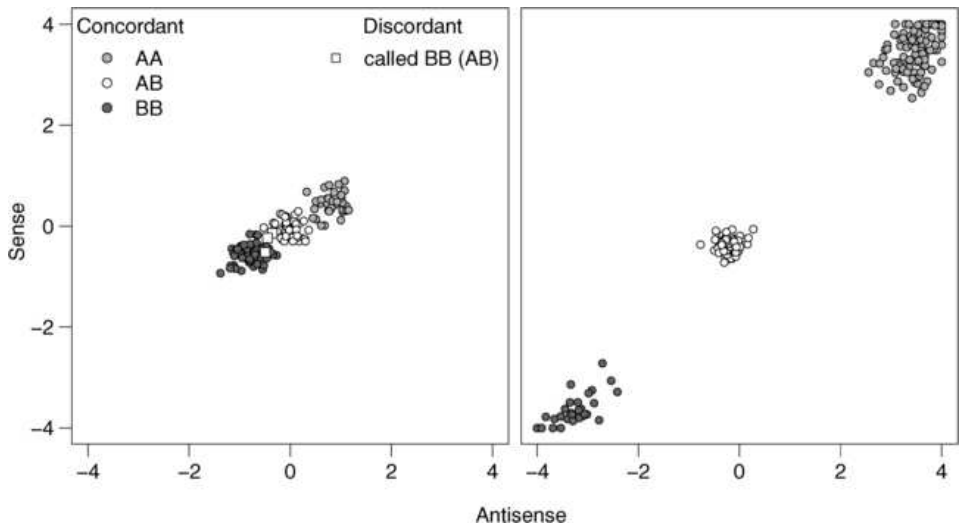

FIG. 1. *HapMap genotype calls (the gold standard) for a bad SNP (left) and a good SNP (right) for 269 samples measured on Affymetrix 100k SNP chips. The HapMap consensus genotype call (taken to be the gold standard) is indicated by color: AA (medium grey), AB (white), and BB (dark grey). The separation between genotype clusters is SNP-specific. This figure motivates an approach that incorporates uncertainty estimates to control smoothing.*

Copy number estimates and genotype calls, however, can provide complementary information. For example, without copy number information, genotype calls alone would not allow for a distinction of LOH due to deletion or uniparental isodisomy (iUPD), which occurs when a subject inherits the same copy of a chromosome (or parts thereof) twice from one parent. While this has been recognized and concurrent analyses have been reported [see, e.g., Zhou et al. (2004, 2005) and Ninomiya et al. (2006)], these analyses were carried out separately for genotype calls and copy number estimates, and the results visually compared. Not until very recently has the need for an integrated analysis of copy number and genotype been addressed for the first time. Colella et al. (2007) propose a Bayesian hidden Markov model approach (QuantiSNP), using both genotype and copy number estimates to infer underlying states (deletions, amplifications, copy neutral regions of homozygosity, etc.) of interest. We caution though that data derived from cancer samples might create substantial problems for HMM based methods like QuantiSNP and our approach: DNA copy numbers larger than three are quite possible in such settings, and thus, the number of possible states expands dramatically. Further, noninteger copy numbers do make sense in tumors due to the mix of normal and abnormal cells in the sample [i.e., mosaicism; see Ting et al. (2006) for an example]. In these settings, copy number based segmentation approaches might be more promising [Olshen et al. (2004), Picard et al. (2005), Venkatraman and Olshen (2007)], in particular, as the definition of a "genotype" is unclear. In this manuscript, we propose a hidden Markov model for the integrated analysis of copy number and genotype estimates, most applicable for abnormalities as a consequence of germline events. We also develop the methodology to integrate genotype and copy number estimate uncertainty measures, and illustrate how integrating such confidence scores of the SNP-level summaries in the HMM can improve inference for the underlying hidden states using simulated and experimental data. These ideas are implemented in the R package *VanillaICE*.

**2. Methods.** In this section we describe three HMMs, dependent on whether genotype estimates (abbreviated $\widehat{\mathrm{GT}}$), copy number estimates (abbreviated $\widehat{\mathrm{CN}}$), or both $\widehat{\mathrm{GT}}$ and $\widehat{\mathrm{CN}}$ are available as defined by three classes of objects for SNP array data [Scharf et al. (2007)].

2.1. *Genotype calls.* Most algorithms that provide SNP-level summaries of genotype assume a copy number of two, and report the genotype estimates as such. We therefore assume throughout this paper that the $\widehat{\mathrm{GT}}$ are of the generic form `AA` or `BB` and `AB` corresponding to $\widehat{\mathrm{HOM}}$ and $\widehat{\mathrm{HET}}$, respectively. The vanilla HMM with hidden states retention ($) and loss (!) of heterozygosity require specification of the initial state probability distribution, the emission probabilities (denoted by $\beta$ below), and the transition probabilities (denoted by $\tau$ below) between the true states. Commonly employed in the literature for the transition probability is the "instability-selection" model for LOH analysis [Newton et al. (1998), Beroukhim et al. (2006)] that describes the dependencies between the underlying states of adjacent SNPs as a function of distance. For any two adjacent SNPs, $\theta$ is defined as the probability that the state of the first marker is not informative (denoted by $I^c$) for the state of the second marker. As the distance between SNPs affects this probability, it is modeled as $\theta(d) = 1 - e^{-2d}$, where $d$ is a genetic or physical distance [e.g., 100 Mb units; see Beroukhim et al. (2006)] between adjacent SNPs. We assume that with probability $1 - \theta(d)$, $\mathrm{SNP}_{(i)}$ is informative (denoted by $I$) for $\mathrm{SNP}_{(i+1)}$ and that no change in state occurs between the adjacent SNPs. For example, this leads to

$$
\begin{aligned}
\tau_{!/!}(d) &= P(!_{i+1}|!_i, d) \\
&= P(!_{i+1}, I|!_i, d) + P(!_{i+1}, I^c|!_i, d) \\
&= P(!_{i+1}|I, !_i, d) \times P(I|!_i, d) \\
&\quad + P(!_{i+1}|I^c, !_i, d) \times P(I^c|!_i, d) \\
&= 1 - \theta(d) + P(!) \times \theta(d),
\end{aligned}
\tag{2.1}
$$

as the probability that the state of $\mathrm{SNP}_{(i+1)}$ is !, given that the state of $\mathrm{SNP}_{(i)}$ with distance $d$ was !. Also,

$$
\tau_{\$/!}(d) = P(\$_{i+1}|!_i, d) = 1 - P(!_{i+1}|!_i, d) = \theta(d) \times P(\$).
\tag{2.2}
$$

$P(\$)$ and $P(!)$ refer to the initial probabilities for $\$$ and $!$, respectively. These initial probabilities can be set as fixed constants using knowledge from previous experiments, or alternatively, learned via the EM algorithm [Dempster, Laird and Rubin (1977)].

Emission probabilities for states $!$ and $\$$ are estimated as

$$(2.3) \quad \beta_!(\widehat{\text{GT}}) \sim \text{Binomial}(p = 0.99) \quad \text{and} \quad \beta_\$(\widehat{\text{GT}}) \sim \text{Binomial}(p = 0.7),$$

where $p$ is the probability of a homozygous genotype call. We use the above probabilities as defaults to reflect values typically seen in experimental data. In a region of retention $\$$, about 70% of SNPs on average are homozygous, while in a region of loss $!$ all SNPs are homozygous, but genotyping errors do occur. Alternatively, as these probabilities are affected by the quality of the assay, they can also be learned via the EM algorithm. In practice, we find that our approaches are rather insensitive to changes in these parameters. It is certainly also possible to use SNP-specific homozygosity rates here if they are known from a reference population. Efficient computation of the probability of the observed sequence given the model is carried out using the forward algorithm as described in Rabiner (1989). The most probable state sequence given the model is calculated via the Viterbi algorithm [Viterbi (1967), Rabiner (1989)].

*Integrating confidence estimates* (*ICE*). When confidence estimates are available, the observed data at a SNP is the genotype call ($\widehat{\text{GT}}$) and the uncertainty measure $S_{\widehat{\text{GT}}}$. The joint distribution of $\widehat{\text{GT}}$ and $S_{\widehat{\text{GT}}}$ depends on the underlying state. For example, if the state for a particular SNP is $!$, the emission probability is

$$(2.4) \qquad \beta_!\{\widehat{\text{GT}}, S_{\widehat{\text{GT}}}\} = f\{\widehat{\text{GT}} \mid !\} \times f\{S_{\widehat{\text{GT}}} \mid \widehat{\text{GT}}, !\}.$$

Note that the first of the two terms on the right-hand side of equation (2.4) is simply the emission probability when estimates of uncertainty are not available. The second term can be understood as a weight for the former term that depends on the confidence with which the call is made. The second term can be approximated using a density estimate of the $S_{\widehat{\text{GT}}}$ where the gold standard is available. For example, using CRLMM on the 269 HapMap samples, the distributions of the respective uncertainty measures for all four possible combinations of called and true genotypes measured on the Affymetrix 100k SNP chips are known. We use kernel based density estimates to obtain the distributions of the confidence scores, given the true and called genotype (separately for the Xba and Hind 50k chips):

$$(2.5) \qquad \begin{aligned} &f\{S_{\widehat{\text{HOM}}} \mid \widehat{\text{HOM}}, \text{HOM}\}, f\{S_{\widehat{\text{HOM}}} \mid \widehat{\text{HOM}}, \text{HET}\}, \\ &f\{S_{\widehat{\text{HET}}} \mid \widehat{\text{HET}}, \text{HOM}\}, f\{S_{\widehat{\text{HET}}} \mid \widehat{\text{HET}}, \text{HET}\}. \end{aligned}$$



The first term in (2.5), for example, denotes the density of the scores when the genotype is correctly called homozygous ($\widehat{\mathrm{HOM}}$) and the true genotype is homozygous (HOM). If the underlying state is !, then the true genotype is always HOM and we assume that

$$
\begin{align}
(2.6) \qquad f\{\mathrm{S}_{\widehat{\mathrm{HOM}}} \mid \widehat{\mathrm{HOM}}, !\} &= f\{\mathrm{S}_{\widehat{\mathrm{HOM}}} \mid \widehat{\mathrm{HOM}}, \mathrm{HOM}\} \quad \text{and} \\
f\{\mathrm{S}_{\widehat{\mathrm{HET}}} \mid \widehat{\mathrm{HET}}, !\} &= f\{\mathrm{S}_{\widehat{\mathrm{HET}}} \mid \widehat{\mathrm{HET}}, \mathrm{HOM}\}.
\end{align}
$$

If the underlying state is \$, then the true genotype can be HET or HOM. We therefore estimate the emission probabilities for state \$ as

$$
\begin{align}
(2.7) \qquad & \beta_{\$}\{\widehat{\mathrm{GT}}, \mathrm{S}_{\widehat{\mathrm{GT}}}\} \\
&= f\{\widehat{\mathrm{GT}} \mid \$\} f\{\mathrm{S}_{\widehat{\mathrm{GT}}} \mid \widehat{\mathrm{GT}}, \$\} \\
&= f\{\widehat{\mathrm{GT}} \mid \$\}(f\{\mathrm{S}_{\widehat{\mathrm{GT}}}, \mathrm{HOM} \mid \widehat{\mathrm{GT}}, \$\} + f\{\mathrm{S}_{\widehat{\mathrm{GT}}}, \mathrm{HET} \mid \widehat{\mathrm{GT}}, \$\}) \\
&= f\{\widehat{\mathrm{GT}} \mid \$\}(f\{\mathrm{S}_{\widehat{\mathrm{GT}}} \mid \mathrm{HOM}, \widehat{\mathrm{GT}}, \$\} f\{\mathrm{HOM} \mid \widehat{\mathrm{GT}}, \$\} \\
&\qquad + f\{\mathrm{S}_{\widehat{\mathrm{GT}}} \mid \mathrm{HET}, \widehat{\mathrm{GT}}, \$\} f\{\mathrm{HET} \mid \widehat{\mathrm{GT}}, \$\}) \\
&= f\{\widehat{\mathrm{GT}} \mid \$\}(f\{\mathrm{S}_{\widehat{\mathrm{GT}}} \mid \mathrm{HOM}, \widehat{\mathrm{GT}}\} f\{\mathrm{HOM} \mid \widehat{\mathrm{GT}}, \$\} \\
&\qquad + f\{\mathrm{S}_{\widehat{\mathrm{GT}}} \mid \mathrm{HET}, \widehat{\mathrm{GT}}\} f\{\mathrm{HET} \mid \widehat{\mathrm{GT}}, \$\}).
\end{align}
$$

The unknown terms in equation (2.7), $f\{\mathrm{HOM} \mid \widehat{\mathrm{GT}}, \$\}$ and $f\{\mathrm{HET} \mid \widehat{\mathrm{GT}}, \$\}$, are also estimated from the HapMap samples.

2.2. *Copy number.* The hidden states for autosomal copy numbers are hemizygous deletion ($\searrow$), two copies ($\rightarrow$), and more than two copies ($\nearrow$). A typical, and from practical experience, quite reasonable assumption when only copy number is considered (applied to aCGH and SNP chip data) is that the logarithm of the copy number estimate, after normalization, is roughly normally distributed around the true log copy number [see, e.g., Zhao et al. (2004)], although slightly heavier tails may also be observed in practice. More important however is the fact that the variability is not necessarily constant across SNPs, which we will address in the ICE HMM. If the variance was assumed to be constant (as done in the vanilla HMM), this parameter can be learned via the EM algorithm [Dempster, Laird and Rubin (1977)], or estimated in a robust manner, for example, using quantiles from the observed data. In the examples presented here, we obtained a robust estimate for the standard deviation of copy number estimates using the 16th and 84th percentiles of the $\log_2$ transformed $\widehat{\mathrm{CN}}$ (corresponding to plus minus one standard deviation from the median). For a state $\mathcal{S}$, the mean $\mu_{\mathcal{S}}$ and



variance $\sigma_S^2$ of the Gaussians used to describe the emission probabilities can be fixed at starting values, or updated by EM. In the vanilla HMM we assume a constant $\sigma^2$ and estimate the emission probabilities for state $\searrow$, for instance (on the $\log_2$ scale, *not* divided by 2), as

$$(2.8) \qquad \beta_{\searrow}(\widehat{CN}) \equiv f(\widehat{CN}|\searrow) \sim N(\mu_S = 0, \text{var} = \sigma^2).$$

The transition probability for the copy number HMM is the same as the one described above.

*Integrating confidence estimates (ICE).*  The emission probabilities for the HMM retains the same location parameters for the Gaussian, but with SNP-specific standard errors for the $\widehat{CN}$. For a given SNP, the emission probability for copy number two ($\rightarrow$), for example, is

$$(2.9) \qquad \beta_{\rightarrow}\{\widehat{CN}|S_{\widehat{CN}}\} \sim N(1, (\sigma \times S_{\widehat{CN}})^2).$$

The scalar $\sigma$ can be estimated from the sample at hand, or set equal to one if $S_{\widehat{CN}}$ measures the actual variability of the copy number estimate around the true copy number.

2.3.  *Copy number and genotype.*  For the joint analysis of copy number and genotype, we extend the transition probabilities in equations (2.1) and (2.2) to the hidden states normal (s), amplification (-), LOH (&), and deletion ()). For the emission probabilities, we assume conditional independence between the copy number estimates and the genotype calls:

$$(2.10) \qquad f(\widehat{CN}, \widehat{GT}|\mathcal{S}) = f(\widehat{CN}|\mathcal{S}) \times f(\widehat{GT}|\mathcal{S}).$$

This equation can be further simplified, as the copy number distribution only depends on the true copy number, and the genotype distribution only depends on the true underlying state being \$ or !. For example, for the deletion state we have

$$(2.11) \quad f\{\widehat{CN}, \widehat{GT} \,|\, )\} = f\{\widehat{CN} \,|\, )\} \times f\{\widehat{GT} \,|\, )\} = f\{\widehat{CN} \,|\, \searrow\} \times f\{\widehat{GT} \,|\, !\}.$$

The terms in equation (2.11) can be estimated as described above for genotype and copy number. Emission probabilities for the other states can be obtained similarly.

2.4.  *Simulation.*  The simulated data are available in the Bioconductor package *VanillaICE*. The simulation comprises one subject's genotype, copy number, and confidences scores for 9165 SNPs on chromosome 1. A description of the 5 features simulated in chromosome 1, referred to by regions A–E, and the underlying hidden states in these regions follows.

*Genotype calls.* With the exception of Regions A, B and C in Figure [2], we simulated 9165 genotypes (the approximate number of SNPs in the two 50k SNP chips) from a Bernoulli distribution with probability 0.7 of $\widehat{\text{HOM}}$. Unless otherwise indicated, confidence scores for $\widehat{\text{GT}}$ were obtained by random draws of confidence scores in the Hapmap data when the CRLMM $\widehat{\text{GT}}$ agreed with the gold-standard as defined by consensus of the HapMap genotyping centers. The reference distributions were made separately for the Affymetrix 50k Xba and Hind chips, and hence, the confidence score sampled for each SNP were made respective to the chip.

*Copy number.* The Affymetrix CNAT tool (version 3.0) was used to obtain $\widehat{\text{CN}}$ for the 9165 SNPs from a presumably normal individual in the HapMap dataset (sample NA06993). Deletions and amplifications were simulated from Gaussian distributions with location parameters $\log_2(1)$ and $\log_2(3)$, respectively. For the scale parameter, we used a robust estimate of the $\log_2$ transformed copy number standard deviation, denoted by $\epsilon$. To illustrate how a confidence score such as a standard error of the copy number estimate could be useful, we simulated standard errors from a shifted Gamma: $\Gamma(1,2) + 0.3$, where 1 is the shape parameter and 2 is the rate parameter. To ascertain the effect of qualitatively high confidence scores on the ICE HMM, we scaled $\epsilon$ by $\frac{1}{2}$. Similarly, to simulate less precise $\widehat{\text{CN}}$, we scaled $\epsilon$ by 2.

Regions A–E were simulated as follows:

- Region A contains 200 SNPs spanning a physical distance of approximately 5 Mb. Two chromosomal segments of 99 homozygous genotypes are separated by a chromosomal segment of 14 kb containing two heterozygous SNPs. Using a 2-state hidden Markov model and using only the simulated genotypes as the observed data, the true underlying states (number of SNPs) are ! (99), $ (2), and ! (99) for the 3 segments, respectively. We augment the genotype calls with copy number estimates obtained directly from the CNAT analysis of a normal Hapmap subject's chromosome 1. Using the 3-state HMM for copy number, the true underlying state is → (200). Modeled jointly, the true underlying state is & (99), s (2) and & (99).

- Region B contains 100 SNPs spanning a physical distance of approximately 2 Mb. Two chromosomal segments each containing 49 SNPs are both in regions of a hemizygous deletion. We assigned a homozygous genotype call to all 98 SNPs in the two hemizygous deletions. The two hemizygous deletions are separated by a chromosomal segment of 360 basepairs with copy number two. To simulate an incorrect genotype call (the true genotype is homozygous for the 2 SNPs on the diploid segment), confidence scores for the two heterozygous SNPs are drawn from the distribution of confidence scores when the CRLMM genotype call of HET was incorrect. Copy number estimates and corresponding confidence scores (standard errors) for the hemizygous deletion were simulated as described above, with the exception that high confidence scores were assigned to the two SNPs in the chromosomal segment with normal copy number. The true underlying state for the genotypes in Region B is ! (100). The true state for the copy number in region B is $\searrow$ (49), $\rightarrow$ (2), and $\searrow$ (49). Modeled jointly, the true states are ) (49), s (2) and ) (49).

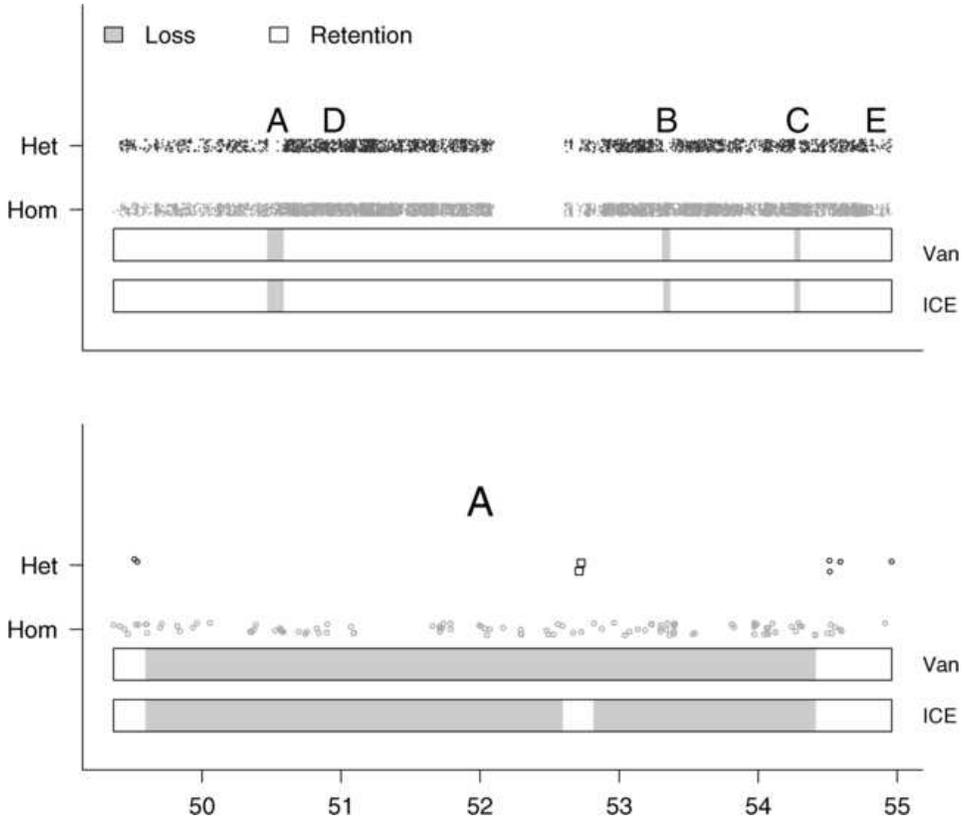

Fig. 2. *A simulated chromosome with 9165 SNPs. Top: The simulated* $\widehat{\mathrm{GT}}$ *with uniform noise added to reduce overplotting (vertical axis) plotted against physical position (horizontal axis). Bottom: A magnification of region A. Two SNPs in region A with high simulated confidence scores are indicated by the square plotting symbol. Regions A–E are described in more detail in Section 2.4. In truth, there are 4 different segments in state loss (!, indicated in light grey above). The predicted hidden states from the vanilla (Van) and ICE HMMs are denoted by color in the two bars beneath the data points. The ICE HMM detects each of the 4 ! segments, whereas the vanilla HMM smoothes over a segment in A containing two heterozygous SNPs at position 52.8 Mb. Utilizing confidence scores for the genotype predictions, the ICE HMM may provide more precise locations for ! breakpoints.*

- Region C is a segment containing 100 homozygous SNPs spanning $< 2$ Mb in a hemizygous deletion. The true underlying states are ! (100) in the genotype HMM, s (100) in the copy number HMM, and ) (100) in the joint HMM.
- Region D contains two segments with copy number 3 ($< 1$ Mb), separated by a diploid segment containing 2 SNPs (9.8 kb). The two amplified fragments are $< 1$ Mb. The true underlying states are \$ (200) in the genotype HMM; ↗ (99), → (2), and ↗ (99) in the copy number HMM; and - (99), s (2) and - (99) in the joint HMM.
- Region E contains a microdeletion spanning 5 SNPs (94 kb) and a microamplification containing 3 SNPs (294 kb). We assigned high confidence scores to the copy number estimates in both regions. The true underlying states are ! (5) and \$ (3) in the genotype HMM, ↘ (5) and ↗ (3) in the copy number HMM and ) (5) and - (3) in the joint HMM.

**3. Results.** This section describes results obtained from fitting HMMs to simulated and experimental data. The HMMs are written in the statistical language R (http://www.r-project.org) using S4 classes and methods [Chambers (1998)]. In particular, the HMM is dependent on whether genotype estimates (abbreviated $\widehat{\text{GT}}$), copy number estimates (abbreviated $\widehat{\text{CN}}$), or both $\widehat{\text{GT}}$ and $\widehat{\text{CN}}$ are available as defined by three classes of objects for SNP array data [Scharpf et al. (2007)]. Organizing the statistical methods in this way allows more flexibility to users interested only in characterizing chromosomal abnormalities in genotype (loss of heterozygosity, LOH) or copy number (deletion or amplification) respectively. When both $\widehat{\text{GT}}$ and $\widehat{\text{CN}}$ are available, the HMM will distinguish between copy-neutral LOH and deletion-induced LOH. We use the term LOH in this context as an unusually long stretch of homozygous SNPs, though these regions can be completely naturally occurring, for example, due to evolutionary pressure on chromosomal segments. For the simulation, we simulate $\widehat{\text{GT}}$ and $\widehat{\text{CN}}$ as described in Section 2.4, analyzing the $\widehat{\text{GT}}$ and $\widehat{\text{CN}}$ separately and then jointly. For the experimental data, we use a HapMap sample with a previously identified region of uniparental isodisomy, a mechanism for copy neutral LOH. Both the simulation and experimental data are based on 100k Affymetrix SNP chips (comprised of the Xba and Hind 50k chips). All figures shown are also available in color as supplementary material at http://biostat.jhsph.edu/~iruczins/publications/sm/.

3.1. *Simulated data.* SNP-level summaries were obtained using a combination of real (experimental) and simulated data for 1965 SNPs measured on chromosome 1 of the 50k Hind and Xba Affymetrix SNP chips, as described in Section 2.4 for additional details. Because the states of the HMM are determined by whether genotype estimates ($\widehat{GT}$), copy number estimates ($\widehat{CN}$), or both $\widehat{GT}$ and $\widehat{CN}$ are available, we organize the results accordingly. For each example, we plot both the predictions of a HMM that uses only the observed SNP-level summaries as input (*vanilla*), and a HMM that integrates confidence estimates (*ICE*) for the SNP-level summaries.

*Genotype HMM.* The hidden states for the genotype HMM are retention (\$) and loss (!) of heterozygosity. In the upper panel of Figure 2 the simulated $\widehat{GT}$ are plotted with uniform noise added to reduce overplotting. The predicted states from the vanilla and ICE HMMs are also shown. The predictions from the vanilla HMM are the same as the predictions of the ICE HMM shown, with the exception of the region (A) magnified in the lower panel of Figure 2, where the ICE HMM correctly identifies the \$ segment. Both approaches miss the 5 SNP spanning microdeletion in region E, but otherwise correctly predict the true underlying states (see Section 2.4 for details). In general, for both the vanilla and ICE HMMs, the Viterbi algorithm (conditional on other parameters of the HMM model) chooses an optimal sequence of states that maximizes the likelihood of the observed genotype calls. The predicted states reflect a trade-off between the likelihood of the observed genotypes given the underlying states, and the transition probabilities. Unlike the vanilla HMM, emission probabilities in the ICE HMM are a function of the confidence scores (as described in Section 2), and factor into the likelihood. Intuitively, a high confidence score at a particular SNP has the effect of giving more weight to the emission probability and less weight to the state of the neighboring SNPs when determining the optimal sequence of states in the Viterbi algorithm. Hence, the sequence of states that maximizes the likelihood of the observed genotype calls differ in the ICE and vanilla HMMs when the confidence scores shifts the balance between the opposing forces of the emission and transition probabilities. In particular, the high confidence scores at the two heterozygous SNPs in region A favor the emission probability for \$, causing two breakpoints in this region of ! and, hence, a more local smoothing of the HMM. Although the emission probability for state \$ is greater than for state ! at these two SNPs in the vanilla HMM, the probability of having two breakpoints in a region of ! for SNPs that are physically close is small as reflected in the transition probability. Therefore, the vanilla HMM provides a smoothing that is less localized, corresponding to a sequence of ! predictions in region A without transitions to the normal state.

*Copy number HMM.* The hidden states for autosomal copy numbers are hemizygous deletion ($\searrow$), normal (two) copies ($\rightarrow$), and more than two copies ($\nearrow$). Figure 3 (upper panel) shows the $\widehat{CN}$ of the simulated dataset. In our simulation, chromosome 1 contains three amplifications $\nearrow$ (two segments in D separated by a segment with normal copy number, and one in

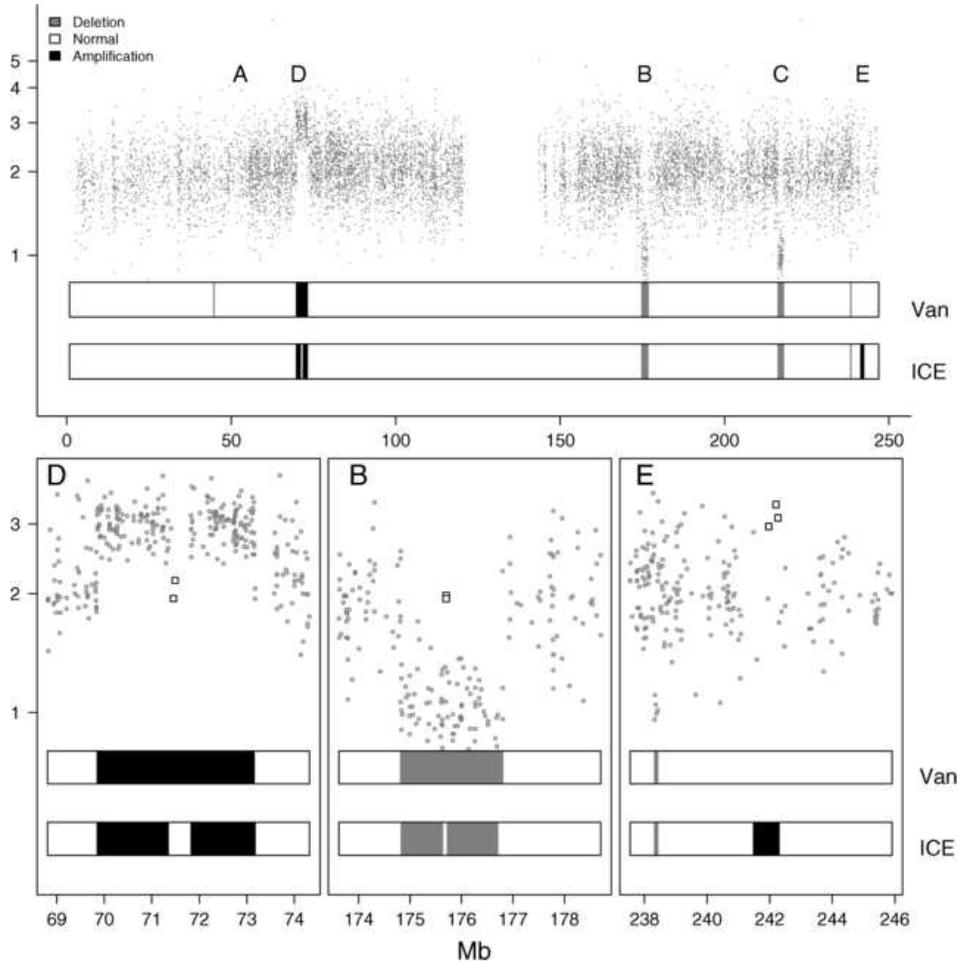

FIG. 3.    *Top: Copy number estimates (vertical axis) versus physical position (horizontal axis) for 9165 SNPs on a simulated chromosome. Bottom: A magnification of regions D, B, and E. High confidence scores for the copy number estimates were simulated for the square points in regions D, B, and E. The two bars beneath the data points in each figure show the predicted hidden states from the vanilla (Van) and ICE HMMs. Note that where the predictions differ in regions D, B, and E, the ICE correctly classified the hidden states. Note that the vanilla HMM also indicates a (spurious) deletion to the left of region A, not indicated by the ICE HMM due to high variability in those copy number estimates.*

E), and four deletions ↘ (two segments in B separated by a segment with normal copy number, and one segment each in regions C and E). Also shown are the predicted states from the vanilla and ICE HMMs, respectively. The predictions from the two HMMs differ in regions B, D, and E magnified in the lower panel. Without confidence estimates for the copy number, the transition probabilities dominate the likelihood as specified by the emission probabilities, and the vanilla HMM smoothes over the two SNPs with copy number 2 in regions B and D, and the amplification in region E. The high confidence scores used in this simulation for the copy number estimates in these regions make the transition between states more favorable, and thus, the ICE HMM makes the transition back to the normal state for regions B and D, and detects the amplification in region E. Note that when the confidences scores for the $\widehat{CN}$ are low, as for the 2 SNPs with copy number near two in the hemizygous deletion in region C, the predictions with ICE and vanilla are identical. Also, the vanilla HMM detects a spurious deletion to the left of region A. As the confidence scores for those copy number estimates were low, the likelihood specified in the ICE HMM does not favor a transition to a nonnormal state.

*Genotype and copy number HMM.* We plot both the $\widehat{GT}$ and $\widehat{CN}$ in the upper panel of Figure 4. By modeling $\widehat{GT}$ and $\widehat{CN}$ simultaneously, we expand the state space of the HMM to include deletion-induced LOH )), copy neutral LOH (&), normal (s), and amplification (-). The predicted states from the vanilla and ICE HMMs are also shown, and differences in predictions are indicated in the lower panel. As before, ICE correctly classifies all SNPs into the respective states, while the vanilla HMM, in the absence of uncertainty estimates, smoothes over some loci (regions A, B, D), and fails to detect the amplification (with high confidence scores) in region E. In contrast, the vanilla HMM does detect the microdeletion in region E. The ability of the vanilla HMM to detect the microdeletion in this example even in the absence of confidence scores is attributable to the additional information that the genotype provides: SNPs in deleted regions all appear as homozygous, in contrast to amplifications, where homozygous and heterozygous SNPs occur. Additionally, the extra genotype information may reduce the occurrence of predicted deletions that are spurious. For instance, in the absence of information on genotype calls in Figure 3, the vanilla HMM predicts a small deletion to the left of region A. As heterozygous genotype estimates in this region are incompatible with a deletion, the vanilla HMM no longer predicts this region to be a deletion in Figure 4.

3.2. *Experimental data.* To illustrate the HMM approaches on experimental data, we used a HapMap sample with a previously identified (but not experimentally confirmed) UPD in chromosome 2. The Affymetrix tool CNAT (version 3.0) and the R software CRLMM were used to obtain SNP-level summaries of copy number and genotype respectively. We caution that at this point in time the $\widehat{GT}$ obtained using CRLMM (or the Affymetrix tools) implicitly assume that the copy number is two—ideally, allele specific estimates should be used, and methods are under development (Rafael

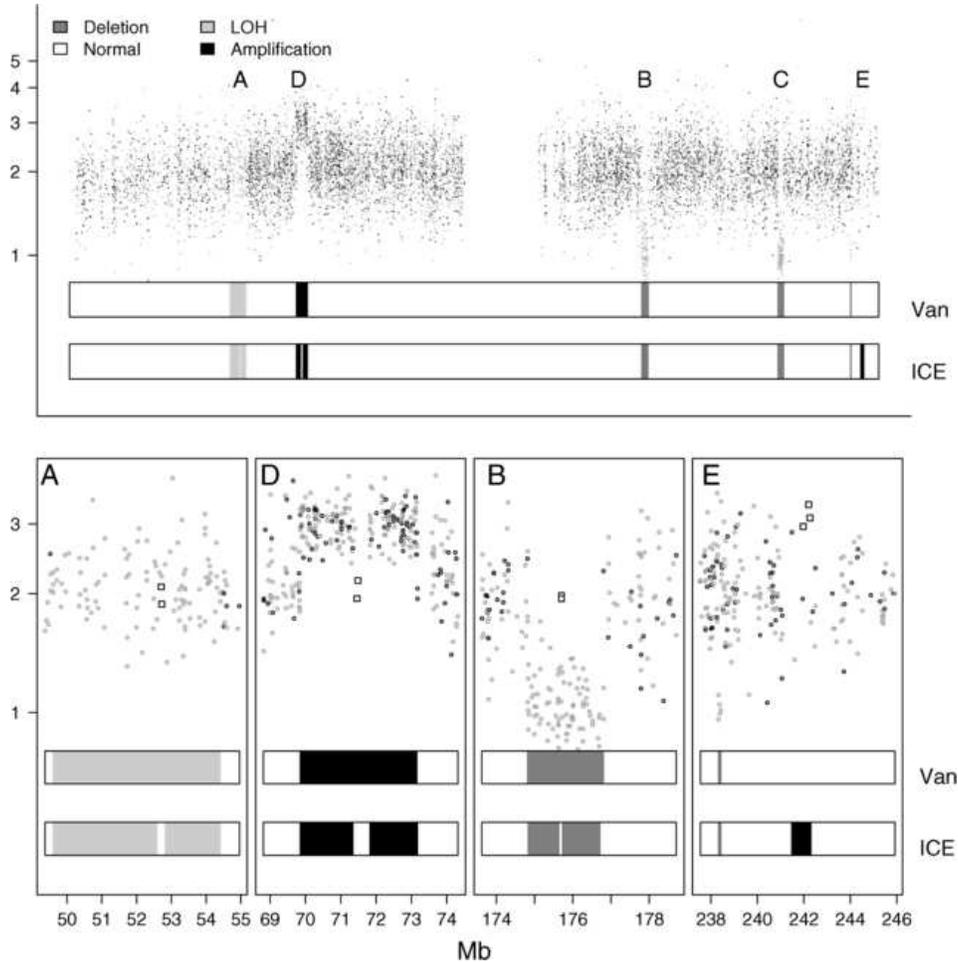

Fig. 4. *Top: The $\widehat{\mathrm{CN}}$ in Figure 3 are superimposed on the $\widehat{\mathrm{GT}}$ in Figure 2. We fit HMMs to the joint observation sequence of $\widehat{\mathrm{CN}}$ and $\widehat{\mathrm{GT}}$ without (vanilla) and with (ICE) confidence scores of the SNP-level summaries. The predictions from these two HMMs are represented by different shades of grey in the two bars beneath the data points in each panel. We used square plotting symbols to indicate SNPs for which we assigned high confidence scores to the genotype and copy number estimates.*

Irizarry, personal communication). Also, software to obtain confidence scores for $\widehat{\mathrm{CN}}$ based on probe-level variability and signal-to-noise ratio on the chip [such as described in Wang et al. (2007)] is not yet available. However, differences in the SNP-specific standard deviations of the $\widehat{\mathrm{CN}}$ across a reference set of 90 HapMap samples have previously been reported [see, e.g., Zhao et al. (2004)], and can be used in a straightforward manner as measures of uncertainty [specifically, using those deviations as the $\mathrm{S}_{\widehat{\mathrm{CN}}}$ in equation (2.9), and

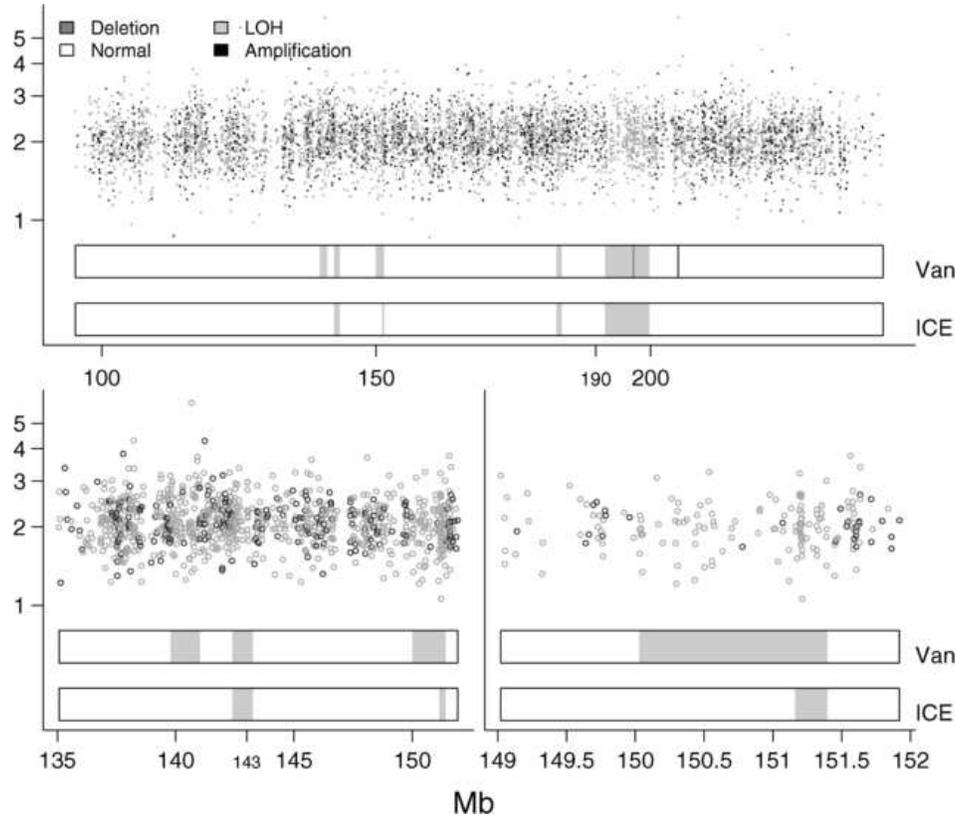

F<small>IG</small>. 5. *Top: A confirmed UPD between 190 and 200 Mb is detected by both HMMs in a HapMap sample from the CEPH dataset. Note that the vanilla HMM incorrectly predicts a small deletion of 3 SNPs in the middle of this region, whereas the ICE HMM provides a more global smoothing of the copy number estimates. Bottom left: a magnified view of three possible LOH regions (not confirmed). Only the middle region (143 Mb) is identified by both HMMs as LOH. Because the CRLMM genotype calls agree with the HapMap consensus, the chromosomal segment containing the two heterozygous SNPs at 140 Mb is not a region of LOH, as predicted by the vanilla HMM. Bottom right: magnification of the vanilla (top) and ICE (bottom) predictions for the feature at 150 Mb. Again, the true genotype calls are heterozygous, and so the ICE HMM correctly identifies the chromosomal segment containing the two heterozygous SNPs as normal.*

estimating the scalar $\sigma$ from the autosomal SNP copy number estimates in the sample].

The upper panel in Figure 5 shows $\widehat{\text{CN}}$ on the vertical axis against physical position on chromosome 2. The region of predominantly called homozygous SNPs at 190–200 Mb is a previously identified UPD [Ting et al. (2006)]. Also shown are the predictions from the vanilla and ICE HMMs. The confirmed UPD between 190 and 200 Mb is detected by both HMMs, though the vanilla HMM incorrectly predicts a small deletion of 3 SNPs in the middle of this region, whereas the ICE HMM provides a more global (and correct) smoothing of the copy number estimates. Also, the vanilla HMM finds a spurious amplification at about 210 Mb. The lower panel on the left provides a magnified view of the region between 135 and 155 Mb, where the vanilla and ICE HMMs differ. Only the middle region (at about 143 Mb) is identified by both HMMs as LOH (we again stress that we use the term LOH here as copy neutral stretches of homozygous SNPs, naturally occurring possibly due to evolutionary pressure on this chromosomal segment). The chromosomal segment at about 140 Mb contains the two heterozygous SNPs (confirmed in the HapMap data, and called as such by CRLMM), and thus is not a region of LOH, as predicted by the vanilla HMM. The lower panel on the right further zooms in on the vanilla and ICE predictions in the region around 150 Mb. The two SNPs with heterozygous genotype calls at about 151 Mb are truly heterozygous SNPs, and therefore, the ICE HMM correctly identifies the chromosomal segment containing these two heterozygous SNPs as normal. Due to the abundance of markers in the segment around 151.25 Mb exclusively called homozygous, the ICE HMM still indicates an LOH segment. Several studies have recognized the abundance of short, copy-neutral, entirely homozygous regions [see, e.g., Beroukhim et al. (2006)]. To illustrate the prevalence of short, homozygous sequences, we fit the vanilla and ICE HMMs to the chromosome 2 data of the 30 CEPH trio parents available from HapMap (60 independent samples), and highlight these copy-neutral, all homozygous regions in Figure 6. Clearly visible is the abundance of these regions, and the enriched locations along chromosome 2 (possibly explained by evolutionary pressure).

3.3. *A vanilla/ICE comparison.* We performed additional simulations to contrast the performances of the vanilla and ICE HMMs. Since large deletions and amplifications can easily be picked up by both approaches, we focused on small deletions and amplifications, spanning between 2 and 10 consecutive SNPs. Since the results were as expected, we only describe the effects of the copy number variability and confidence scores on the detection of small deletions in detail.

The experimental data consisted of genotype calls and copy number as described in Section 2.4. Copy number confidence scores were obtained by weighting the robust estimate of the within-chip $\log_2$ copy number standard deviation by the standardized SNP specific standard deviation derived from a reference set of 90 HapMap samples (e.g., this weight for one particular SNP was the ratio of the across sample standard deviation for the SNP and the median of all those numbers across all SNPs). Simulated in these data were 450 sets of copy number estimates and confidence scores for deletions ranging from two to ten consecutive SNPs (50 data sets for each deletion size). The locations of the deletions were randomly selected on chromosome 1 for each data set. The copy numbers in the deletions were simulated from a log-normal distribution with mean zero (indicating a true DNA copy number equal to one), and a standard deviation equal to a scaled version of the SNP specific variability described above. The scalar $K$ controlled whether more ($K < 1$) or less ($K > 1$) precise copy number data than average were encountered in the deletion. For both vanilla and ICE, we calculated for each simulated data set the difference in log likelihoods between making a transition to the state for deletion ()) from the normal state (s) for the range of the simulated deletion (and back after the deletion), versus staying in the normal state throughout. In other words, we calculated the difference of the log likelihood of the true state sequence minus the log likelihood of assigning the normal state s to all SNPs.

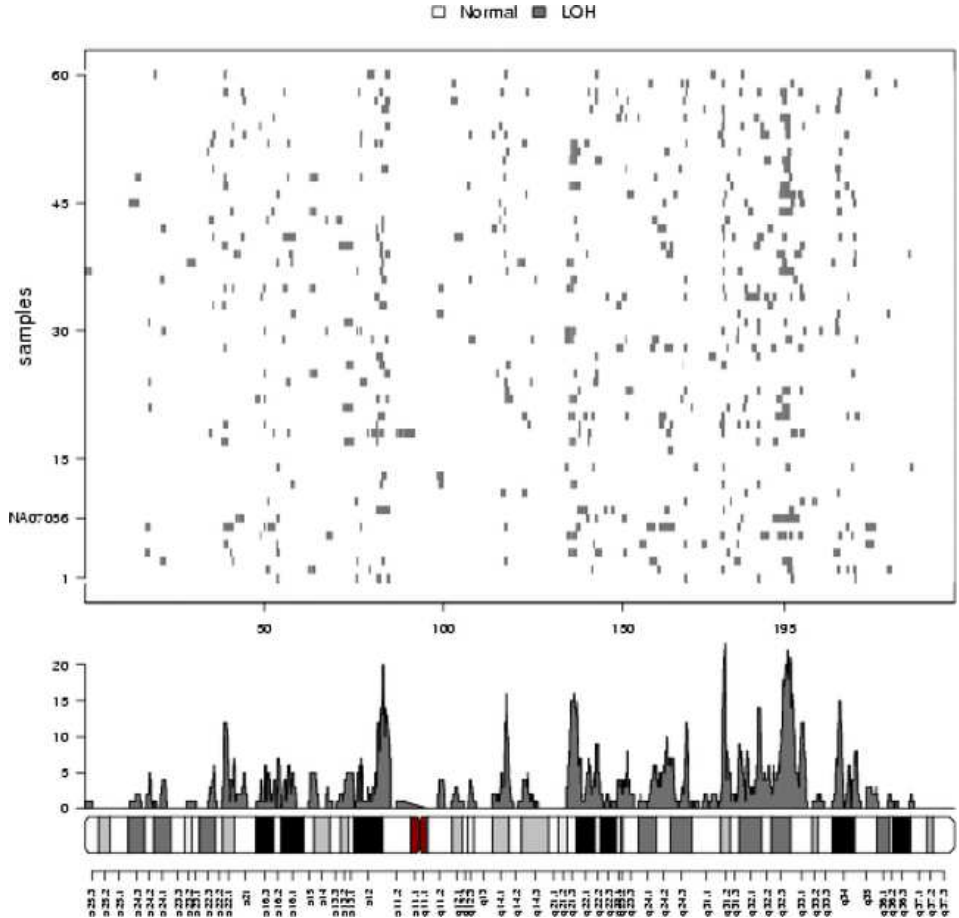

FIG. 6.  *An image of the predictions from the vanilla HMM fit to chromosome 2 of the 60 parental samples in the CEPH trios dataset (top). The x-and y-coordinates used for the image are physical position and subject, respectively. Subject NA07056 has a confirmed UPD at 195 Mb. Also plotted are the frequencies of LOH across the 60 samples (middle) and the cytoband (bottom).*

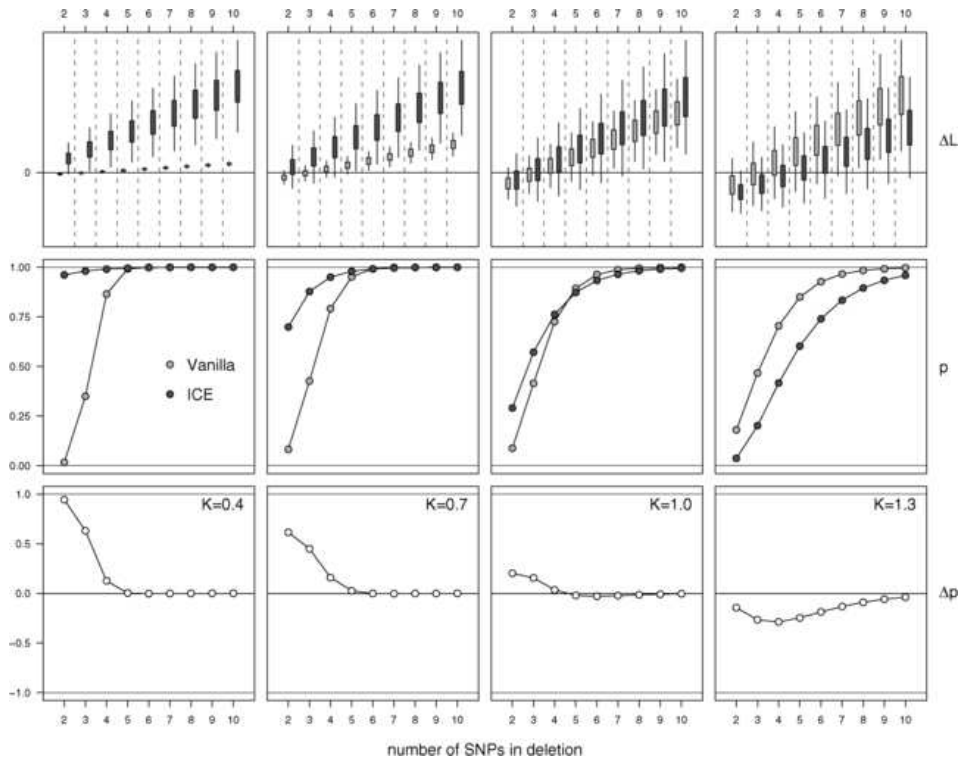

FIG. 7.    *Differences between the log likelihoods for the correct and incorrect state sequences for the vanilla (light grey) and ICE (dark grey) HMMs are indicated in the upper panels. The differences are shown for deletions of different sizes (horizontal axis), and four different scale parameters $K$ for the copy number estimate variability in the simulated deletions (0.4, 0.7, 1.0, 1.3, left to right). The data were scaled to fit the panels, and slightly smoothed from the raw data by exploiting an obvious mean and variance relationship. The middle row of panels shows the estimated probabilities of the differences in log likelihoods being positive (e.g., the proportion of instances when the correct model was favored over the incorrect one), assuming normality of the differences in the log likelihoods. The lower row of panels shows the estimated differences in these probabilities between ICE and vanilla.*

The upper row of panels in Figure 7 indicate the distributions of the differences in the log likelihoods for both the vanilla (light grey) and ICE (dark grey) HMMs, shown for the deletions of different sizes, and using four different scale parameters $K$. For the first two panels, the variability in the simulated copy number estimates in the deleted region was less than in the original data (the standard deviations were reduced to 40% and 70% of the original, respectively), and for the fourth panel the standard deviation in the simulated copy number estimates in the deleted region was increased by 30%. The middle row of panels shows the respective estimated probabilities of the differences in log likelihoods being positive, for example, the proportion of instances when the correct model was favored over the incorrect one. The lower row of panels shows the difference in these probabilities between ICE and vanilla. Quite obvious is the fact that the ability to detect microdeletions of a few SNPs depends on precise data, and the knowledge of that precision. For example, when the standard deviation of the simulated copy number estimates in the deletion was reduced to 40%, ICE was able to consistently detect even the smallest deletions, while vanilla was only able to do so for deletions of size 5 or larger (left panels). Naturally, larger deletions are easier to detect for both methods. As the quality of the data decreases (simulated here as an increase in the variability of the copy number estimates in the deletion), the ability of ICE to detect the deletion suffers substantially, while vanilla is almost agnostic to these changes. When the standard deviation of the simulated copy number estimates in the deletion was increased by 30%, vanilla picked up the deletion more often than ICE (right panels). The reason for this is as follows: since the variability in the copy number estimates is increased, the evidence of a deletion being present decreases, and ICE acknowledges this fact by incorporating the confidence estimates. Thus, the decrease in the proportion of instances where ICE favors a deletion over the normal state is a feature of the algorithm. The price to pay, otherwise, is in the number of false positives (i.e., the number of incorrectly inferred deletions at other loci). Simulating 200 "synthetic" normal chromosome 1q arms with $K = 1.3$ across all SNPs, vanilla indicated spurious small deletions in 50 of these artificial chromosomal arms (for a total of 86 incorrect state predictions), while ICE indicated none.

**4. Discussion.** Chromosomal DNA varies between individuals at the level of entire chromosomes, chromosomal segments, and changes in small genomic regions down to one nucleotide (including single nucleotide polymorphisms, SNPs). Many of these variations appear to be completely benign, but some are known or suspected to be associated with disease. Association studies often use some SNPs (in candidate gene studies) or hundreds of thousands of SNPs (in genome wide association studies) as potential candidates or markers of genes to investigate the relationship between genotype and phenotype. However, the abundance of copy number variations in the human genome and their role in disease have played an increasingly prominent role. In particular, the "common disease, common variant" paradigm has been challenged for some diseases [McClellan, Susser and King (2007); see, e.g., Sebat et al. (2007) for a case study on autistic and apparently normal subjects]. Undoubtedly, this change is due in part to the recent technological advancement, in particular, on high density single nucleotide polymorphism (SNP) microarrays which allow for the detection of these alterations. Besides copy number variations such as deletions and duplications, copy-neutral stretches of homozygosity can also be of scientific interest, as uniparental disomy as one such example has been implicated in disease.

Copy number variations and loss of heterozygosity can arise through somatic and germline events. In this manuscript, we developed methods most applicable for abnormalities as a consequence of germline events. Undoubtedly, the stochastic process as defined by our transition probability could be too rigid for the analysis of data arising from a cancer sample, where microdeletions as well as a loss of an entire chromosomal arm might be present. Further, noninteger copy numbers do make sense in such samples due to the mix of normal and abnormal cells in the sample [i.e., mosaicism; see Ting et al. (2006) for an example], while we assume the copy numbers to be integers in our approach. While rare, noninteger copy numbers may occur even in "normal" genomes (this can occur throughout the body or in specific regions), and thus, may pose a problem for our algorithm. In general, even if our method could be extended to allow for noninteger copy numbers (at least the HMM for copy numbers, since the definition of "genotype" is unclear in such a setting), the ability to pick up noninteger copy numbers obviously depended on the quality of the data, the length of the non-normal region, and the actual value of said copy number. For example, delineating a small mosaic region in a sample with 95% normal cells and 5% of cells with a hemizygous deletion would likely not be possible.

Our paper builds on a modular approach for analyzing SNP chip data, extending the functionality of statistical algorithms that pre-process probe-level data to produce SNP-level summaries of genotype and copy number. Noticeably, these approaches have mostly been developed for the Affymetrix platform (such as CRLMM for improved genotype estimates), but our ideas are portable to other high throughput platforms such as Illumina. In particular, the vanilla HMM only relies on genotype ($\overline{\text{CN}}$) and copy number ($\overline{\text{GT}}$) estimates without any confidence scores, which can be exported directly from the Beadstudio software (http://www.illumina.com/). With one noticeable (and very recent) exception suggested by Colella et al. (2007), previous approaches using HMMs have considered genotype and copy number separately, not simultaneously in a single unifying statistical model that allows for the detection of copy number changes as well as copy neutral stretches of homozygosity in the genome. In this sense, this manuscript is not the first to propose such a unifying approach, albeit ours differs in several aspects from the Bayesian HMM of Colella et al. (2007). In particular, the incorporation of uncertainty estimates can be critical, for example, in the detection of microdeletions. The investigation of one particular sample as discussed in this manuscript, however, does not allow for conclusive statements how the detected alterations are associated with the phenotype. In particular, it has been well established that copy number variations and copy neutral stretches of homozygous genotypes are prevalent in many phenotypically normal individuals. Identifying features that may be associated with a particular phenotype are better handled by statistical models for between-sample variation in studies with phenotypically normal and diseased populations. Such models reside in the next tier of our modular approach to the analysis of SNP chip data and are an extension of the ideas presented here.

In summary, we developed a HMM for SNP chips using the joint observation sequence of copy number ($\widehat{CN}$) and genotype ($\widehat{GT}$) estimates as input. We demonstrated that a HMM model that uses both $\widehat{CN}$ and $\widehat{GT}$ can, for example, distinguish copy-neutral LOH from deletion-induced LOH. We also demonstrated how pre-processing algorithms that provide confidence scores of SNP-level summaries can be integrated into the emission probabilities of the HMM to control smoothing in a probabilistic framework, and showed that this can lead to much improved results. Specifically, confidence estimates allow smoothing to be more local or global depending on the uncertainty of the pointwise estimates. We demonstrated how high confidence scores helped in identifying a very small amplification otherwise missed (Figure 4, region E), while low confidence scores for $\widehat{CN}$ and $\widehat{GT}$ had the desirable effect of providing a more global smoothing (Figure 5). In particular, in the experimental data example, this helped to reduce the number of regions identified as LOH in the vanilla HMM, and eliminated the (presumably, spurious) indication of a small deletion and a small amplification. We believe that the ability to detect microdeletions and microamplifications could be of utmost importance to explain the genetic basis of many diseases. Undoubtedly, this ability will greatly depend not only on the number of markers investigated (such as the number of SNPs used on a particular platform) and the quality of the data produced (i.e., the precision of the genotype and copy number estimates), but also on how the uncertainty of the estimates is utilized. In this sense, we hope that our method and software provides a useful tool for the scientific community.

**Acknowledgments.**  We gratefully acknowledge the support from and helpful discussions with Benilton Carvalho, Rafael Irizarry, and the members of the Pevsner Laboratory.

## REFERENCES

AFFYMETRIX (2006). Brlmm: An improved genotype calling method for the genechip human mapping 500k array set. Technical report, Affymetrix, Inc.

AGGARWAL, A., LEONG, S. H., LEE, C., KON, O. L. and TAN, P. (2005). Wavelet transformations of tumor expression profiles reveals a pervasive genome-wide imprinting of aneuploidy on the cancer transcriptome. *Cancer Res.* **65** 186–194.

AGUIRRE, A. J., BRENNAN, C., BAILEY, G., SINHA, R., FENG, B., LEO, C., ZHANG, Y., ZHANG, J., GANS, J. D., BARDEESY, N., CAUWELS, C., CORDON-CARDO, C., REDSTON, M. S., DEPINHO, R. A. and CHIN, L. (2004). High-resolution characterization of the pancreatic adenocarcinoma genome. *Proc. Natl. Acad. Sci. USA* **101** 9067–9072.

ALTUG-TEBER, O., DUFKE, A., POTHS, S., MAU-HOLZMANN, U. A., BASTEPE, M., COLLEAUX, L., CORMIER-DAIRE, V., EGGERMANN, T., GILLESSEN-KAESBACH, G., BONIN, M. and RIESS, O. (2005). A rapid microarray based whole genome analysis for detection of uniparental disomy. **26** 153–159.

BEROUKHIM, R., LIN, M., PARK, Y., HAO, K., ZHAO, X., GARRAWAY, L. A., FOX, E. A., HOCHBERG, E. P., MELLINGHOFF, I. K., HOFER, M. D., DESCAZEAUD, A., RUBIN, M. A., MEYERSON, M., WONG, W. H., SELLERS, W. R. and LI, C. (2006). Inferring loss-of-heterozygosity from unpaired tumors using high-density oligonucleotide SNP arrays. *PLoS Comput. Biol.* **2** e41.

CARVALHO, B., BENGTSSON, H., SPEED, T. P. and IRIZARRY, R. A. (2007). Exploration, normalization, and genotype calls of high-density oligonucleotide SNP array data. *Biostatistics* **8** 485–499.

CHAMBERS, J. M. (1998). *Programming with Data.* Springer, New York.

COLELLA, S., YAU, C., TAYLOR, J. M., MIRZA, G., BUTLER, H., CLOUSTON, P., BASSETT, A. S., SELLER, A., HOLMES, C. C. and RAGOUSSIS, J. (2007). QuantiSNP: An objective Bayes hidden-Markov model to detect and accurately map copy number variation using SNP genotyping data. *Nucleic Acids Res.* **35** 2013–2025.

DEMPSTER, A., LAIRD, D. and RUBIN, D. (1977). Maximum likelihood from incomplete data via the EM algorithm. *J. Roy. Statist. Soc. Ser. B* **39** 1–38. MR0501537

DI, X., MATSUZAKI, H., WEBSTER, T. A., HUBBELL, E., LIU, G., DONG, S., BARTELL, D., HUANG, J., CHILES, R., YANG, G., MEI SHEN, M., KULP, D., KENNEDY, G. C., MEI, R., JONES, K. W. and CAWLEY, S. (2005). Dynamic model based algorithms for screening and genotyping over 100 K SNPs on oligonucleotide microarrays. *Bioinformatics* **21** 1958–1963.

DUTT, A. and BEROUKHIM, R. (2007). Single nucleotide polymorphism array analysis of cancer. *Curr. Opin. Oncol.* **19** 43–49.

EICHLER, E. E., NICKERSON, D. A., ALTSHULER, D., BOWCOCK, A. M., BROOKS, L. D., CARTER, N. P., CHURCH, D. M., FELSENFELD, A., GUYER, M., LEE, C., LUPSKI, J. R., MULLIKIN, J. C., PRITCHARD, J. K., SEBAT, J., SHERRY, S. T., SMITH, D., VALLE, D. and WATERSTON, R. H. (2007). Completing the map of human genetic variation. *Nature* **447** 161–165.

EILERS, P. H. C. and DE MENEZES, R. X. (2005). Quantile smoothing of array CGH data. *Bioinformatics* **21** 1146–1153.

ENGEL, E. (2006). A fascination with chromosome rescue in uniparental disomy: Mendelian recessive outlaws and imprinting copyrights infringements. *Eur. J. Hum. Genet.* **14** 1158–1169.

FREEMAN, J. L., PERRY, G. H., FEUK, L., REDON, R., MCCARROLL, S. A., ALTSHULER, D. M., ABURATANI, H., JONES, K. W., TYLER-SMITH, C., HURLES, M. E., CARTER, N. P., SCHERER, S. W. and LEE, C. (2006). Copy number variation: New insights in genome diversity. *Genome Res.* **16** 949–961.

FRIDLYAND, J., SNIJDERS, A., PINKEL, D., ALBERTSON, D. and JAIN, A. (2004). Hidden Markov models approach to the analysis of array CGH data. *J. Multivariate Anal.* **90** 132–153. MR2064939

GUHA, S., LI, Y. and NEUBERG, D. (2006). *Bayesian Hidden Markov Modeling of Array CGH Data.* Berkeley Electronic Press.

HOUSEMAN, E. A., COULL, B. A. and BETENSKY, R. A. (2006). Feature-specific penalized latent class analysis for genomic data. *Biometrics* **62** 1062–1070. MR2297677

HSU, L., SELF, S. G., GROVE, D., RANDOLPH, T., WANG, K., DELROW, J. J., LOO, L. and PORTER, P. (2005). Denoising array-based comparative genomic hybridization data using wavelets. *Biostatistics* **6** 211–226.

HUA, J., CRAIG, D. W., BRUN, M., WEBSTER, J., ZISMANN, V., TEMBE, W., JOSHIPURA, K., HUENTELMAN, M. J., DOUGHERTY, E. R. and STEPHAN, D. A. (2007). SNiPer-HD: Improved genotype calling accuracy by an expectation-maximization algorithm for high-density SNP arrays. *Bioinformatics* **23** 57–63.

HUANG, J., WEI, W., CHEN, J., ZHANG, J., LIU, G., DI, X., MEI, R., ISHIKAWA, S., ABURATANI, H., JONES, K. W. and SHAPERO, M. H. (2006). CARAT: A novel method for allelic detection of DNA copy number changes using high density oligonucleotide arrays. *BMC Bioinformatics* **7** 83.

HUANG, T., WU, B., LIZARDI, P. and ZHAO, H. (2005). Detection of DNA copy number alterations using penalized least squares regression. *Bioinformatics* **21** 3811–3817.

HUPE, P., STRANSKY, N., THIERY, J. P., RADVANYI, F. and BARILLOT, E. (2004). Analysis of array CGH data: From signal ratio to gain and loss of DNA regions. *Bioinformatics* **20** 3413–3422.

KENNEDY, G. C., MATSUZAKI, H., DONG, S., MIN LIU, W., HUANG, J., LIU, G., SU, X., CAO, M., CHEN, W., ZHANG, J., LIU, W., YANG, G., DI, X., RYDER, T., HE, Z., SURTI, U., PHILLIPS, M. S., BOYCE-JACINO, M. T., FODOR, S. P. A. and JONES, K. W. (2003). Large-scale genotyping of complex DNA. *Nat. Biotechnol.* **21** 1233–1237.

LAFRAMBOISE, T., HARRINGTON, D. and WEIR, B. A. (2006). PLASQ: A generalized linear model-based procedure to determine allelic dosage in cancer cells from SNP array data. *Biostatistics* **8** 323–326.

LAI, W. R., JOHNSON, M. D., KUCHERLAPATI, R. and PARK, P. J. (2005). Comparative analysis of algorithms for identifying amplifications and deletions in array CGH data. *Bioinformatics* **21** 3763–3770.

LAI, Y. and ZHAO, H. (2005). A statistical method to detect chromosomal regions with DNA copy number alterations using SNP-array-based CGH data. *Comput. Biol. Chem.* **29** 47–54.

LIN, M., WEI, L. J., SELLERS, W. R., LIEBERFARB, M., WONG, W. H. and LI, C. (2004). dChipSNP: Significance curve and clustering of SNP-array-based loss-of-heterozygosity data. *Bioinformatics* **20** 1233–1240.

MCCLELLAN, J. M., SUSSER, E. and KING, M. C. (2007). Schizophrenia: A common disease caused by multiple rare alleles. *Br. J. Psychiatry* **190** 194–199.

NANNYA, Y., SANADA, M., NAKAZAKI, K., HOSOYA, N., WANG, L., HANGAISHI, A., KUROKAWA, M., CHIBA, S., BAILEY, D. K., KENNEDY, G. C. and OGAWA, S. (2005). A robust algorithm for copy number detection using high-density oligonucleotide single nucleotide polymorphism genotyping arrays. *Cancer Res.* **65** 6071–6079.

NEWTON, M. A., GOULD, M. N., REZNIKOFF, C. A. and HAAG, J. D. (1998). On the statistical analysis of allelic-loss data. *Stat. Med.* **17** 1425–1445.

NINOMIYA, H., NOMURA, K., SATOH, Y., OKUMURA, S., NAKAGAWA, K., FUJIWARA, M., TSUCHIYA, E. and ISHIKAWA, Y. (2006). Genetic instability in lung cancer: Concurrent analysis of chromosomal, mini- and microsatellite instability and loss of heterozygosity. *Br. J. Cancer* **94** 1485–1491.

OLSHEN, A. B., VENKATRAMAN, E. S., LUCITO, R. and WIGLER, M. (2004). Circular binary segmentation for the analysis of array-based DNA copy number data. *Biostatistics* **5** 557–572.

PICARD, F., ROBIN, S., LAVIELLE, M., VAISSE, C. and DAUDIN, J. J. (2005). A statistical approach for array CGH data analysis. *BMC Bioinformatics* **6** 1471–2105.

RABBEE, N. and SPEED, T. P. (2006). A genotype calling algorithm for affymetrix SNP arrays. *Bioinformatics* **22** 7–12.

RABINER, L. R. (1989). A tutorial on hidden Markov models and selected applications in speech recognition. *Proc. IEEE* **77** 257–286.

REDON, R., ISHIKAWA, S., FITCH, K. R., FEUK, L., PERRY, G. H., ANDREWS, T. D., FIEGLER, H., SHAPERO, M. H., CARSON, A. R., CHEN, W., CHO, E. K., DALLAIRE, S., FREEMAN, J. L., GONZALEZ, J. R., GRATACOS, M., HUANG, J., KALAITZOPOULOS, D., KOMURA, D., MACDONALD, J. R., MARSHALL, C. R., MEI, R., MONTGOMERY, L., NISHIMURA, K., OKAMURA, K., SHEN, F., SOMERVILLE, M. J., TCHINDA, J., VALSESIA, A., WOODWARK, C., YANG, F., ZHANG, J., ZERJAL, T., ZHANG, J., AR-MENGOL, L., CONRAD, D. F., ESTIVILL, X., TYLER-SMITH, C., CARTER, N. P., ABU-RATANI, H., LEE, C., JONES, K. W., SCHERER, S. W. and HURLES, M. E. (2006). Global variation in copy number in the human genome. *Nature* **444** 444–454.

ROBINSON, W. P. (2000). Mechanisms leading to uniparental disomy and their clinical consequences. *Bioessays* **22** 452–459.

SCHARPF, R. B., TING, J. C., PEVSNER, J. and RUCZINSKI, I. (2007). SNPchip: R classes and methods for SNP array data. *Bioinformatics* **23** 627–628.

SEBAT, J., LAKSHMI, B., MALHOTRA, D., TROGE, J., LESE-MARTIN, C., WALSH, T., YAMROM, B., YOON, S., KRASNITZ, A., KENDALL, J., LEOTTA, A., PAI, D., ZHANG, R., LEE, Y. H., HICKS, J., SPENCE, S. J., LEE, A. T., PUURA, K., LEHTIMAKI, T., LEDBETTER, D., GREGERSEN, P. K., BREGMAN, J., SUTCLIFFE, J. S., JOBANPUTRA, V., CHUNG, W., WARBURTON, D., KING, M. C., SKUSE, D., GESCHWIND, D. H., GILLIAM, T. C., YE, K. and WIGLER, M. (2007). Strong association of de novo copy number mutations with autism. *Science* **316** 445–449.

SHAH, S. P., XUAN, X., DELEEUW, R. J., KHOJASTEH, M., LAM, W. L., NG, R. and MURPHY, K. P. (2006). Integrating copy number polymorphisms into array CGH anal-ysis using a robust HMM. *Bioinformatics* **22** e431–e439.

SHAW-SMITH, C., REDON, R., RICKMAN, L., RIO, M., WILLATT, L., FIEGLER, H., FIRTH, H., SANLAVILLE, D., WINTER, R., COLLEAUX, L., BOBROW, M. and CARTER, N. P. (2004). Microarray based comparative genomic hybridisation (array-CGH) detects sub-microscopic chromosomal deletions and duplications in patients with learning disabil-ity/mental retardation and dysmorphic features. *J. Med. Genet.* **41** 241–248.

SZATMARI, P., PATERSON, A. D., ZWAIGENBAUM, L., ROBERTS, W., BRIAN, J., LIU, X. Q., VINCENT, J. B., SKAUG, J. L., THOMPSON, A. P., SENMAN, L., FEUK, L., QIAN, C., BRYSON, S. E., JONES, M. B., MARSHALL, C. R., SCHERER, S. W., VIELAND, V. J., BARTLETT, C., MANGIN, L. V., GOEDKEN, R., SEGRE, A., PERICAK-VANCE, M. A., CUCCARO, M. L., GILBERT, J. R., WRIGHT, H. H., ABRAMSON, R. K., BETAN-cur, C., BOURGERON, T., GILLBERG, C., LEBOYER, M., BUXBAUM, J. D., DAVIS, K. L., HOLLANDER, E., SILVERMAN, J. M., HALLMAYER, J., LOTSPEICH, L., SUTCLIFFE, J. S., HAINES, J. L., FOLSTEIN, S. E., PIVEN, J., WASSINK, T. H., SHEFFIELD, V., GESCHWIND, D. H., BUCAN, M., BROWN, W. T., CANTOR, R. M., CONSTANTINO, J. N., GILLIAM, T. C., HERBERT, M., LAJONCHERE, C., LEDBETTER, D. H., LESE-MARTIN, C., MILLER, J., NELSON, S., SAMANGO-SPROUSE, C. A., SPENCE, S., STATE, M., TANZI, R. E., COON, H., DAWSON, G., DEVLIN, B., ESTES, A., FLODMAN, P., KLEI, L., MCMAHON, W. M., MINSHEW, N., MUNSON, J., KORVATSKA, E., RODIER, P. M., SCHELLENBERG, G. D., SMITH, M., SPENCE, M. A., STODGELL, C., TEPPER, P. G., WIJSMAN, E. M., YU, C. E., ROGE, B., MANTOULAN, C., WITTEMEYER, K., POUSTKA, A., FELDER, B., KLAUCK, S. M., SCHUSTER, C., POUSTKA, F., BOLTE,

S., Feineis-Matthews, S., Herbrecht, E., Schmotzer, G., Tsiantis, J., Papanikolaou, K., Maestrini, E., Bacchelli, E., Blasi, F., Carone, S., Toma, C., Van Engeland, H., de Jonge, M., Kemner, C., Koop, F., Langemeijer, M., Hijmans, C., Staal, W. G., Baird, G., Bolton, P. F., Rutter, M. L., Weisblatt, E., Green, J., Aldred, C., Wilkinson, J. A., Pickles, A., Le Couteur, A., Berney, T., McConachie, H., Bailey, A. J., Francis, K., Honeyman, G., Hutchinson, A., Parr, J. R., Wallace, S., Monaco, A. P., Barnby, G., Kobayashi, K., Lamb, J. A., Sousa, I., Sykes, N., Cook, E. H., Guter, S. J., Leventhal, B. L., Salt, J., Lord, C., Corsello, C., Hus, V., Weeks, D. E., Volkmar, F., Tauber, M., Fombonne, E. and Shih, A. (2007). Mapping autism risk loci using genetic linkage and chromosomal rearrangements. *Nat. Genet.* **39** 319–328.

Ting, J., Ye, Y., Thomas, G., Ruczinski, I. and Pevsner, J. (2006). Analysis and visualization of chromosomal abnormalities in SNP data with SNPscan. *BMC Bioinformatics* **7** 25.

Venkatraman, E. S. and Olshen, A. B. (2007). A faster circular binary segmentation algorithm for the analysis of array CGH data. *Bioinformatics* **23** 657–663.

Viterbi, A. (1967). Error bounds for convolution codes and an asymptotically optimal decoding algorithm. *IEEE Trans. Inform. Theory* **13** 260–269.

Wang, P., Kim, Y., Pollack, J., Narasimhan, B. and Tibshirani, R. (2005). A method for calling gains and losses in array CGH data. *Biostatistics* **6** 45–58.

Wang, W., Carvalho, B., Miller, N., Pevsner, J., Chakravarti, A. and Irizarry, R. A. (2007). Estimating genome-wide copy number using allele specific mixture models. In *RECOMB* 137–150.

Willenbrock, H. and Fridlyand, J. (2005). A comparison study: Applying segmentation to array CGH data for downstream analyses. *Bioinformatics* **21** 4084–4091.

Zhao, X., Li, C., Paez, J. G., Chin, K., Jänne, P. A., Chen, T. H., Girard, L., Minna, J., Christiani, D., Leo, C., Gray, J. W., Sellers, W. R. and Meyerson, M. (2004). An integrated view of copy number and allelic alterations in the cancer genome using single nucleotide polymorphism arrays. *Cancer Res.* **64** 3060–3071.

Zhou, X., Mok, S. C., Chen, Z., Li, Y. and Wong, D. T. W. (2004). Concurrent analysis of loss of heterozygosity (loh) and copy number abnormality (cna) for oral premalignancy progression using the affymetrix 10k SNP mapping array. *Hum. Genet.* **115** 327–330.

Zhou, X., Rao, N. P., Cole, S. W., Mok, S. C., Chen, Z. and Wong, D. T. (2005). Progress in concurrent analysis of loss of heterozygosity and comparative genomic hybridization utilizing high density single nucleotide polymorphism arrays. *Cancer Genet. Cytogenet.* **159** 53–57.

Zlotogora, J. (2004). Parents of children with autosomal recessive diseases are not always carriers of the respective mutant alleles. *Hum. Genet.* **114** 521–526.

R. B. Scharpf
I. Ruczinski
Department of Biostatistics
Johns Hopkins Bloomberg School
    of Public Health
Baltimore, Maryland 21205
USA
E-mail: rscharpf@jhsph.edu
        ingo@jhu.edu

G. Parmigiani
The Sidney Kimmel Comprehensive
    Cancer Center
Johns Hopkins University
Baltimore, Maryland 21205
USA
E-mail: gp@jhu.edu

J. Pevsner
Department of Neurology
Kennedy Krieger Institute
Baltimore, Maryland 21205
USA
E-mail: pevsner@kennedykrieger.org